\documentclass[a4paper]{spie}
\usepackage[dvips]{graphicx}

\title{Far infrared polarimeter with very low instrumental polarization}

\author{E. S. Battistelli, M. DePetris, L. Lamagna, R. Maoli,
F. Melchiorri, E. Palladino, G. Savini \skiplinehalf \supit{} {\it
Department of Physics, University of Rome "La Sapienza", Rome,
Italy} \skiplinehalf \supit{} \skiplinehalf \supit{} P. D.
Mauskopf, A. Orlando \skiplinehalf \supit{} {\it Department of
Physics and Astronomy, University of Wales, Cardiff} }

\authorinfo{Corresponding authors: Battistelli
Elia and Savini Giorgio. E-mail:
elia.stefano.battistelli@roma1.infn.it , savini@roma1.infn.it ,
address: P.le Aldo Moro 2, 00185, Rome, Italy}

\begin{document}
\maketitle

\begin{abstract}

    After a short analysis of the main problems involved in the
construction of a Far Infrared polarimeter with very low
instrumental noise, we describe the instrument that will be
employed at MITO telescope to search for calibration sources and
investigate polarization near the CMB anisotropy peaks in the next
campaign (Winter 2002-03).

\end{abstract}


\section{INTRODUCTION}
\label{sec:intro}

    Since the detection of the CMB anisotropy by COBE \cite{Smoot1992}
and the further characterization of its Doppler peaks by Boomerang
\cite{DeBernardis2000} and Maxima \cite{Hanany2000}, extensive
calculations and simulations have been performed to predict the
degree of polarization of the CMB foreseen by Rees \cite{Rees1968}
that later had been upper limited ($0.1$mK) by Caderni et al.
\cite{Caderni1978}. By introducing the Stokes formalism and the
elements of polarization genesis in the Microwave Background, in
this work we analyze the various contributions to an expected
polarized signal that is to be measured from ground experiments
(N.Gnedin and N.A.Silant'ev 1997\cite{Gnedin1997}). We stress the
importance of accurate removal of instrumental spurious
polarization and propose a possible technique of data analysis to
reduce systematic effects mostly related to atmospheric
contamination in small-beam ($4^\prime\div5^\prime$ ) Far Infrared
ground based experiments\cite{DePetris2001}. We then describe our
MITO-Pol experiment (working at $120-360GHz$) and its forthcoming
upgrade in detector sensitivity, installation at MITO telescope,
and the first light of the instrument with beam calibration before
the planned campaign, during Winter 2002-2003, for measurements of
polarization field near high peaks of CMB anisotropy as predicted
by Arbuzov\cite{Arbuzov1997}.

\section{Searching for CMB polarization}
\label{sec:theory}

    As radiation decouples from matter (redshift $z \sim 1000$ ), the
intensity of photons arriving to us from the Last Scattering
Surface(LSS), after Thomson scattering, is peaked in the direction
normal to the initial propagation, and with the residual
polarization parallel to the incident one\cite{White98}. We see
then that in an isotropic context contributions of polarized
radiation scattered by LSS electrons cancel out two by two,
leaving only quadrupole distributions of intensity as possible
"source" of polarized (linear) radiation.
\begin{figure}[ht]
 \begin{center}
  \begin{tabular}{c}
  \includegraphics[height=4cm]{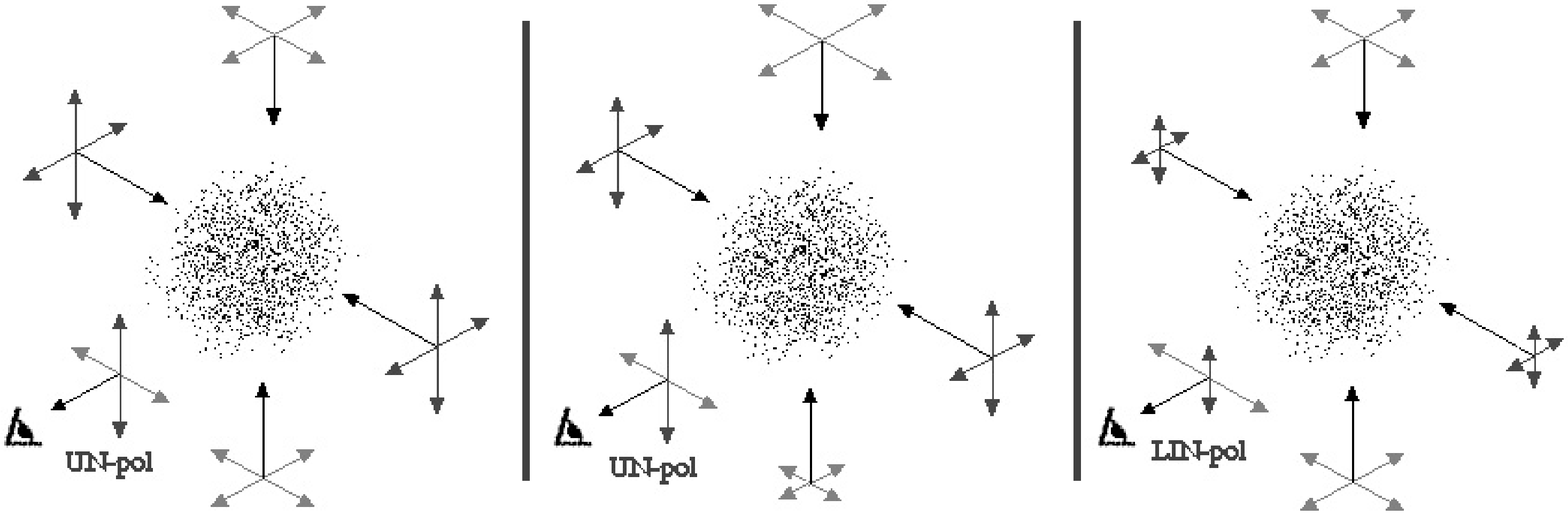}
  \end{tabular}
 \end{center}
 \caption[example]{Isotropic distributions cancel out by 90 degree contributions, while dipole
 intensity cancel out 4 by 4 summing mutually opposed contributions and cancelling with the
 orthogonal ones. A quadrupole anisotropy is necessary to observe a degree of linear polarization.}
\end{figure}

    Theoretical frameworks for the origin of polarized radiation have
been numerous throughout the years. By giving a complete
description of radiative transfer following Boltzmann's equation
\cite{MelcVitt1996}, or by describing in terms of Stokes
Parameters the expected Polarization of the Cosmic Microwave
Background (CMBP)\cite{HuWhite1997}, important cosmological
informations can be obtained. Particularly interesting is the
study of the power spectra of the possible polarization modes,
E-modes and B-modes\cite{Zaldarriaga1997}, and the correlation of
the temperature density spectrum with them. The E-modes spectrum,
essentially deriving from scalar density perturbations on LSS, is
expected to be well correlated with the temperature spectrum while
B-modes spectrum is not. Vector perturbations due to the vortical
motions of the matter, generate mainly B-modes while tensor
perturbations, which can be associated to gravitational waves,
generate both E-modes and B-modes. A detection of the CMBP can be
helpful in order to remove some still existing degeneracies in the
determination of the cosmological parameters. The polarization
spectrum is sensitive to the presence of gravitational waves at a
wider angular scale range than the temperature spectrum.
Furthermore a CMB polarization characterization can tell us
informations on the duration of the recombination era and can give
light on the reionization process which can not be studied solely
by temperature spectrum. The expected polarization level is of the
order of $10\%$ of the anisotropies. Anyway, as Arbuzov et
al.\cite{Arbuzov1997} has predicted, we expect local polarization
maxima of $\sim 30 \div 45 \%$ around $3 \sigma$-peaks of
anisotropies mainly between the cross-levels $ 1.5\sigma$ and $
2.5\sigma$, shaped as rings around the peaks themselves.

    Let us refrain from going deep in all the possible cosmological
scenarios that can be favored by detection of polarization of the
CMB, to concentrate on the upper limits that have been recently
set on this foreground. In table \ref{table1} we have summarized
some of the currently undergoing experiments that should be able
to give further information on microwave sky polarization or to
set more stringent upper limits on the E and B modes power
spectrum.

    Generally experiments can be roughly divided as investigating in
small and large angular scales. At low frequencies (i.e.
$\nu<100GHz$) very stringent upper limits have been posed: $\sim
10 \mu K$ at both large \cite{Keating2002} and small
\cite{Hedman2002} angular scales. Considering the frequency
dependence of the expected CMB polarization compared with the
expected foreground emission (see $\S$\ref{sec:foregrounds}), the
best frequency range to perform CMB polarization measurements
seems to be above $100GHz$. Furthermore, given the perspective of
having a polarized intensity of a peak of E-mode polarization at a
few $\mu$K, we choose to follow a less global approach of sky
coverage in favour of a detailed analysis of temperature
anisotropy peaks present in existing CMB maps.
\begin{table}[h] \caption{Recent and future
experiments of CMB polarization measurement.} \label{table1}
\begin{center}
\begin{tabular}{|c|c|c|c|c|c|c|c|}

\hline \rule[-1ex]{0pt}{3.5ex}  Name & $\nu$/$\Delta \nu $($GHz$)
& Beam & Sensitivity & Coverage & Polarimeter & Upp.limit \\

\hline \rule[-1ex]{0pt}{3.5ex} CBI & $26-36$ &
$3^{\prime}-30^{\prime}$ & $20$ ($\mu$K/night) & - &
Interferometer & - \\

\rule[-1ex]{0pt}{3.5ex} ATCA & $8.7$ & $2^{\prime}$ & - & $6$ pix
& Interferometer &  $\le 11\mu K$ \\

\rule[-1ex]{0pt}{3.5ex}POLATRON & $90/20$ & $2.5^{\prime}$ & $.7$
($m$K$/\sqrt{Hz}$) & 850 pix & OMT & -
\\

\rule[-1ex]{0pt}{3.5ex}PolKa & $350$ &
$10^{\prime\prime}-1^{\prime}$ & - & - & rotating analyzer & $\le
300\mu$K \\

\rule[-1ex]{0pt}{3.5ex}MITO-Pol & $150-350$ & $5^{\prime}$ & $2$
($m$K$ \cdot \sqrt{s}$) & $10^\circ\times 10^\circ$ &
d.F.r.+M.Puplett &
-
\\

\rule[-1ex]{0pt}{3.5ex}PIQUE & $40$,$90$ /$16$ & $14^{\prime}$ &
$2$ ($m$K$\cdot \sqrt{s}$) & $\sim 25$ pix & OMT & $\le 10\mu K$
\\

\rule[-1ex]{0pt}{3.5ex}MilanoPol & $33$/$1.5$ & $7^\circ-14^\circ$
& $1$ ($m$K$/\sqrt{Hz}$) & - & correlator & - \\

\rule[-1ex]{0pt}{3.5ex}MilanoPol2 & $33$/$1.5$ & $15^{\prime}$ &
$1$ ($m$K$/\sqrt{Hz}$) & - & correlator & - \\

\rule[-1ex]{0pt}{3.5ex}POLAR & $26-36$,$90-100$ & $7^\circ$ & $1$
($m$K$/\sqrt{Hz}$) & $1844^\circ$ & OMT & $\le 10\mu$K \\

\rule[-1ex]{0pt}{3.5ex}COMPASS & $26-36$,$90-100$ &
$10^{\prime}$-$20^{\prime}$ & $1$ ($m$K$\cdot \sqrt{s}$) & - & OMT
& - \\ \hline

\rule[-1ex]{0pt}{3.5ex}MAXIPOL & $150$,$240$,$410$ & $10^{\prime}$
& $.041$ ($m$K$ \cdot \sqrt{s}$) & $10^6$ pix & grid.pol.+HWP & -
\\

\rule[-1ex]{0pt}{3.5ex}Boom2K & $90$,$150$,$240$,$410$ &
$10^{\prime}$ & $200$ ($\mu$K$/\sqrt{Hz}$) & $10^6$ pix &
polarization abs. & -
\\

\rule[-1ex]{0pt}{3.5ex}BAR-Sport & $32$,$90$ &
$30^{\prime}$,$12^{\prime}$ & $.5$ , $.7$ ($m$K$\cdot \sqrt{s}$) &
$20^\circ\times20^\circ$ & OMT & -
\\ \hline

\rule[-1ex]{0pt}{3.5ex}MAP & $22$,$30$,$40$,$60$,$90$ &
$13^{\prime}-56^{\prime}$ & $35$ ($\mu$K$/pix$) & full & OMT & -
\\


\rule[-1ex]{0pt}{3.5ex}Planck-HFI & $6$Ch $\in$ $100-857$ &
$5^{\prime}-33^{\prime}$ & $6$ ($\mu$K$/pix$) & full & - & -
\\

\rule[-1ex]{0pt}{3.5ex}SPOrt & $22$,$32$,$60$,$90$/$10$\% &
$7^\circ$ & $1$ ($m$K$\cdot \sqrt{s}$) & $82\%$ & OMT & -
\\ \hline

\end{tabular}
\end{center}
\end{table}

\section{MITO-Pol: a polarimeter with low instrumental polarization for the very far-infrared region}
\label{sec:Fotopola}

    The MITO polarimeter, to be mounted at the MITO telescope on Italian
Alps, employs two bolometric detectors, separated by a polarizing
wire-grid, to perform continuous monitoring of modulated polarized
signals incoming at the focal plane of the telescope after passing
in a double-Fresnel rhomb and a modified Martin-Puplett
interferometer. This composed modulation allows us to extract a
fixed polarized signal supposed to exist "behind" any spurious and
non-constant polarizing foreground.

\subsection{MITO: Millimetre Infrared Testagrigia Observatory}

    Located in the highest reaches of the Alps, at $3480m$ a.s.l., the observatory is
devoted to Far Infrared measurements of cosmological interest. The
contents of precipitable water vapour have been computed by HITRAN
and the observational results confirm a low level in pwv, reaching
peaks of antarctic level for the last months of Winter. The
telescope is a $2.6$m primary Cassegrain in altazimuthal
configuration, the secondary mirror with wobbling capabilities
will remain fixed on-axis while modulation with the polarimeter
will be obtained by a rotating double Fresnel rhomb (see $\S$
\ref{sub:instrument}). Measurement of CMB anisotropy and SZ-effect
\cite{DePetris2001}$^{,}$\cite{Battistelli2002} have already
proven the efficiency and reliability of the telescope.
\begin{figure}[ht]
 \begin{center}
  \begin{tabular}{c}
  \includegraphics[height=7cm]{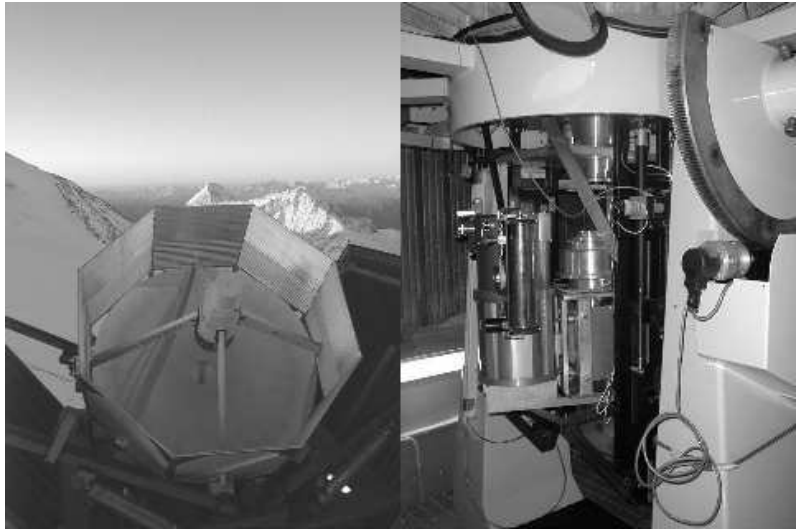}
  \end{tabular}
 \end{center}
 \caption[example]{Left:MITO telescope with ground shields on the Plateau Rosa
   site ($3480$m). Right: Mito-POL cryostat mounted with the interferometer and the
   double-Fresnel rhomb at the focal plane of the telescope.}\label{fig:mito}
\end{figure}

\subsection{The instrument}\label{sub:instrument}

\begin{figure}[t]
 \begin{center}
  \begin{tabular}{c}
  \includegraphics[height=7cm]{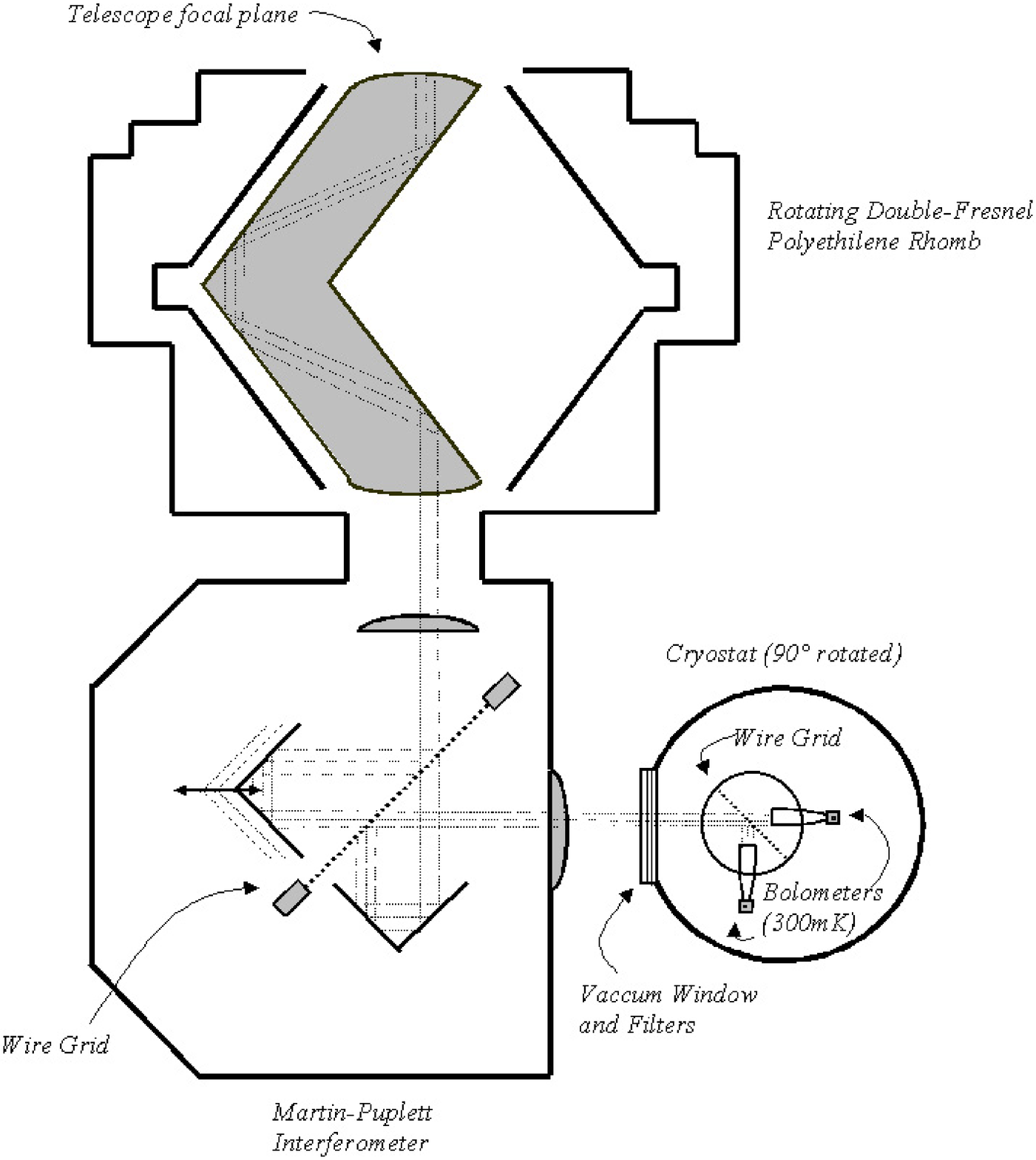}
  \end{tabular}
 \end{center}
 \caption[example]{Section of final instrument setup.}\label{fig:setup}
\end{figure}

    Radiation focused by the telescope enters the rotating double Fresnel-rhomb
that has been adopted to modulate a polarized signal, leaving the
non-polarized background (or foreground) untouched. The double
rhomb is made of high-density polyethilene with a refractive index
selected to obtain total internal reflection of the focused rays
with minimal dispersion. On reflection, linearly polarized signals
are phase-shifted by $\pi/4$ through each rhomb, obtaining the
equivalent of a half-wave retarder by employing two rhombs like in
fig.\ref{fig:setup}. The polarized signal will thus rotate at an
angular speed double than the physical speed $\omega$ of the
double rhomb. If we adopt Stokes parameters, we can express this
element with the following Mueller matrix:

\[ \left( \begin{array}{cccc} 1 & 0 & 0 & 0 \\ 0 & cos 4\omega t & sin 4\omega t & 0 \\
 0 & -sin 4\omega t & cos 4\omega t & 0 \\ 0 & 0 & 0 & -1 \\
\end{array} \right)\]

    On exiting the double-rhomb, radiation enters the Martin-Puplett
interferometer and passes through a $45^{\circ}$ \footnote{with
respect to the interferometer axis.} axis wire-grid that splits
polarized radiation in its two orthogonal components, along with
acting as a beamsplitter of unpolarized radiation. These two
separate components, travel towards two roof-shaped mirrors (see
fig.\ref{fig:setup}), one of which shifts to change the
path-length of the beam introducing a phase shift $\phi =2\pi
l/\lambda$, where $l$ is the increase of the path and $\lambda $
is the wavelength. The two beams recombine and enter the
photometer after being refocused, to be splitted by a final
wire-grid (positioned on the cold flange), at the entrance of the
two radiation collectors (i.e. two $f/4$ Winston cones that
concentrate the radiation on the detectors). If we compose the
Mueller matrices of the optical elements employed, given an
entrance signal $(I,Q,U,V)$ we obtain an output on the detectors:

\begin{equation}
S_{Ch1/Ch2}=\frac{1}{2} \left( \begin{array}{cccc} 1 & \pm 1 & 0 &
0
\\ \pm 1 & 1 & 0 & 0 \\ 0 & 0 & 0 & 0 \\ 0 & 0 & 0 & 0 \\\end{array}
\right) \left( \begin{array}{cccc} 1 & 0 & 0 & 0 \\ 0 & cos \phi &
0 & sin \phi \\ 0 & 0 & -1 & 0 \\ 0 & sin \phi & 0 & -cos \phi
\\ \end{array} \right) \left( \begin{array}{cccc} 1 & 0 & 0 & 0 \\
0 & cos 4\omega t & sin 4\omega t& 0 \\ 0 & sin 4\omega t & -cos
4\omega t & 0 \\ 0 & 0 & 0 & -1 \\ \end{array} \right) \left(
\begin{array}{c} I \\ Q \\ U \\ V \\ \end{array}
\right)
\end{equation}

so as to have on the two channels, separated from the last
wire-grid:

$$ S_{Ch1}= \frac{1}{2}[I+(Qcos4\omega t+Usin4\omega t)cos\phi
-Vsin\phi] $$

$$ S_{Ch2}= \frac{1}{2}[I-(Qcos4\omega t+Usin4\omega t)cos\phi
+Vsin\phi] $$

We see how subtraction of channels, normalized with the sum,
yields a good indication of polarized signals with modulation of Q
and U Stokes parameters at a frequency 4 times that of the
mechanical rotation of the Fresnel-rhombs. This would avoid
spurious polarized signals originating by slight-misalignment of
the rhomb, rotating with $\omega $ frequency, to give an
unpolarized asymmetric contribution.

\begin{figure}[ht]
 \begin{center}
  \begin{tabular}{c}
  \includegraphics[height=3cm]{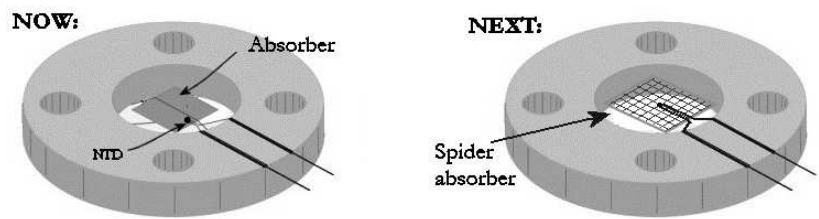}
  \end{tabular}
 \end{center}
 \caption[example]{The detectors employed are composite bolometers with NTD-germanium
 as thermistors.}\label{fig:bolo}
\end{figure}

    The detectors that we are employing at the moment are composite
bolometers with NTD-germanium as thermistors (see
fig.\ref{fig:bolo}). Their operating temperature in the cold
flange of the HDL-8 model Infrared Labs Inc. $^{3}He$ cryostat is
of $300 \div 305m$K reached\footnote{After $^{4}He$ transfer and
thermalization.} with a $\sim 4$ hours procedure, and has a time
duration working cycle of $\sim 18$ hours. The NET of these
detectors is $2m$K$\cdot\sqrt{s}$ in laboratory conditions (higher
background and working temperature). If we consider this minimum
signal as a final operational value, we need an integration time
of about twenty minutes per pixel ($1.7'$ obtained by a beam of
$5'$ $FWHM$) to obtain a S/N ratio greater than $1$ (for signal
amplitude as the one predicted by Arbuzov\cite{Arbuzov1997}). In
future campaigns we will employ spider-web
bolometers\cite{Mauskopf1997} which will reduce NET to BLIP
(Background Limited Infrared Photodetection) conditions allowing
us to compute a better analysis of sources either by obtaining a
higher S/N ratio, or by allowing us to map sources in a
considerably shorter integration time.

    The frequency band selection and the reduction of the background
on the bolometers have been performed employing a filters sequence
anchored at the various thermal shields of the photometer. The
first filter is the z-cut quartz $4mm$ vacuum window which is
characterized by high transmission up to $100cm^{-1}$. At the
$N_{2}$ stage, at $77$K, we have used again a quartz filter with
antireflecting coating obtained by means of diamond powder on a
side and black polyethylene on the other side. Black poly removes
radiation above $400cm^{-1}$ (NIR and visible) while the diamond
powder cuts the UV radiation. Two Yoshinaga filters have been
placed at the $^{4}He$ stage at $1.6$K. These produce an
electro-magnetic cut above $55cm^{-1}$. Two further Yoshinaga
filters (one per winston cone) have been placed at the $^{3}He$
stage at $300m$K just at the Winston cone inputs so to block
possible residual IR radiation that would saturate the bolometers.
The final band selection is performed by means of two embedded
mesh filters placed again at the cone inputs. Mesh filters allow a
very sharp selection with a cut-off frequency
$\nu_{cut-off}=12cm^{-1}$. The low frequency band threshold is
determined by the diameter of the output holes of the Winston
cones which behave as high-pass filters with a cut-on frequency
$\nu_{cut-on}=4cm^{-1}$.

    Signal amplification and bolometer bias
supply have been performed in order to reduce the high detector
impedance, to produce an amplification of the order of $1000$ and
to keep as low as possible the receiver noise. The first
requirement has been satisfied by using a JFET amplifier in a
common drain configuration (characterized by a very high input
impedence) directly installed on the $^{4}He$ flange. The second
requirement has been obtained by means of a operational amplifier
in a non-inverting configuration. A detailed study has been
performed\cite{Pisano2000} in order to reduce the electronic noise
of the amplification system as well as of the bolometer bias
supply. The overall system noise has been estimated and
experimentally confirmed to be of the order of $5nV/\sqrt{Hz}$.

    MitoPol experiment has been installed at MITO telescope during the
last FotoMito campaign for SZ-Effect measurements (Winter
2001-2002), to verify mechanical matching to the telescope,
optical coupling, operational procedures in measurement
conditions, and possible electronic problems in situ. A
preliminary test of the instrument without the double
Fresnel-rhomb has been done to measure the beam of the instrument
as a photometer. The first light allowed us to understand how
important is a good optical alignment as well as the reduction of
electric cross-talk between the channels. This last problem, for
instance, has induced us (also considering the imminent detector
change) to move into differential amplification system. Several
calibrations by chopping two blackbodies ($77$ and $300$K) behind
a polarizer have been performed. Tests and further calibrations
are going on, in particular in order to solve all the problems
connected to the installation of the new detectors.

\section{Instrumental polarization and spurious atmospheric polarization}
\label{sec:instrumental}

    In order to detect faint polarized signals, in the presence of a
much greater unpolarized background, it is necessary to reduce at
minimum level all possible contributions to spurious polarization
effects (which can result in an additional polarized signal or in
reducing a present polarization (i.e. de-polarization)). First of
all, in the selection of the spectral region of observation,
filters employed must be carefully chosen to avoid materials or
components that have oriented structures or that have been
machined with a final preferential axis. Metal meshes and grids
must be avoided, and similarly powder-pressed filter that undergo
baking may retain a global anisotropy in grain-orientation (i.e.
fluorogold). Also, thin polymer films that are "stretched" may
contribute to partial polarization along the tension axis.

    The other important issue that must be considered is polarization (or
de-polarization) induced by oblique reflection from good
conductors as is the case of aluminum alloys (i.e. mirrors). From
this point of view, Winston cones have been shown to be strongly
de-polarizing so that they cannot be used before the splitting of
the two polarizations\cite{Pisano1999}. This overall effect, that
has been modeled precisely in geometric terms for optical
astronomy, can cause spurious polarization as far as $10^{-4}$ for
a non-polarized point source at $1$ arcminute from the optical
axis in a $f/10$ Cassegrain configuration telescope\cite{Sen1997}.
At larger wavelengths the effect decreases due to change in the
complex refractive index of the surface. Renbarger et
al.\cite{Renbarger1998} have practically measured spurious
polarization of this kind for large reflection angles ($15 \div
45^{\circ}$)\footnote{The measurements had the intent of
demonstrating low ($10^{-4}$) spurious polarization for off-axis
telescopes measuring polarized emission by dust, possibly
correlated with galactic magnetic fields.} (for $\lambda \sim 250
\div 1000 \mu m$ ), finding a value of a few tenths of percent for
multiple reflections at $45^{\circ}$ angles. Calculations at much
smaller angles to determine the level of spurious polarization
introduced by a Cassegrain $f/4$ telescope at millimeter
wavelengths have been made with a ray tracing
program\cite{Pisano2000}$^{,}$\cite{Pisano1999} giving a level
just short of $10^{-6}$ for spurious polarization inside a $5$
arcminute beam. Much care must be anyway taken in dealing with
off-axis sources (for instance in drift scan procedure), this is
one of the reasons why we have chosen a $5'$ beam.

    Instrumental spurious polarization can therefore be drastically
reduced by employing on axis telescope configurations and
arranging the optical layout in such a way that the various
channels have exactly the same beam. Unfortunately this symmetry
requirement cannot always be exactly obtained to the required
level. For a ground based experiment one has to take into account
the atmosphere emission that, even if unpolarized, has an
intensity much higher than the signal we want to detect. A
polarization induced by not perfectly symmetric reflections could
thus simulate a polarization signal. On the other hand one of the
main advantage of polarization experiments with respect to
photometric experiments is the possibility to reduce atmospheric
fluctuations by differencing the two polarization states of the
same sky region\cite{Melc1997}. The atmospheric fluctuations
cannot anyway be totally removed by simply subtracting the signals
of the two channels if the two beams are misaligned. This effect
is much larger for point-like sources in comparison with extended
sources which completely fill the beams. The not perfect beams
alignment results in the presence of non-gaussian atmospheric
fluctuations, observed at "large" $\sigma$. These can be
eliminated in data analysis by filtering the signal fluctuations
(beyond $n \sigma$) with an appropriate filter (with $n$ to be
found by maximizing the signal to noise ratio). A similar analysis
has been applied to Fotomito SZ data and has shown best values for
$n$ of a few units giving the highest signal to noise ratios.


\section{Foregrounds and systematics in the detection of CMBP}
\label{sec:foregrounds}

    As for the search of CMB anisotropies in the far-infrared region,
polarization measurements require a detailed knowledge of
foregrounds and how to remove them. The advantage of having some
non-polarized foregrounds, is leveled by the lack of information
regarding the expected polarized fraction of single contributions.
Apart from the atmosphere, other important foregrounds that must
be taken into account are dust contamination due to emission of
aligned small particles and grains in primordial or galactic
magnetic fields \cite{Draine1999}, galactic free-free emission
that can give a $10\%$ contribution concentrated in HII regions
(due to Thomson scattering), and synchrotron emission, both
galactic and extra-galactic, the former of course depending on
galactic latitude, dominating at low frequencies but with a
significant contamination at mm wavelengths \cite{Lubin1981}. Some
of these foregrounds can be efficiently removed if measurements at
different frequencies are provided, others (as in the case of
polarized dust emission) can be compared to consistent patterns of
galactic magnetic field when mapping, so that these can be ignored
in the presence of expected anisotropy ring patterns inconsistent
with magnetic field lines if present \cite{Kogut2000}. Thus, by
performing observations in sky regions which are particularly
clean from foregrounds, studying the detected polarized signals
and the dependence on the frequency and on the sky coordinates,
foregrounds can be drastically reduced and finally eliminated from
the CMB polarization measurements.
\begin{figure}[ht]
 \begin{center}
  \begin{tabular}{c}
  \includegraphics[height=7cm]{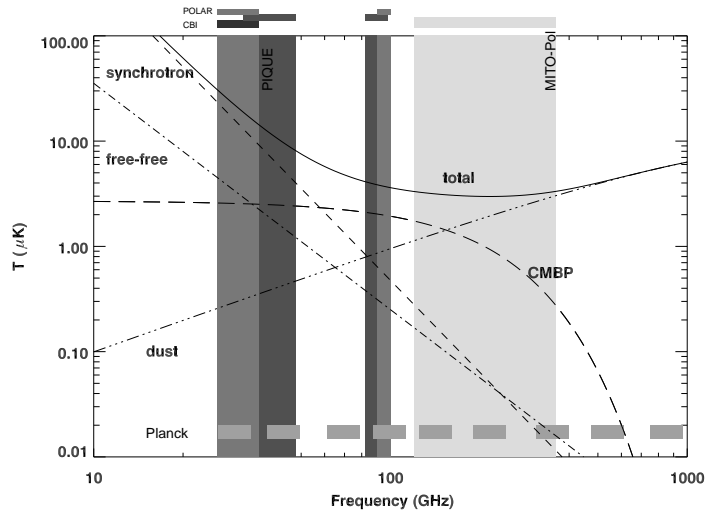}
  \end{tabular}
 \end{center}
 \caption[example]{Expected polarized foregrounds, with corresponding frequency bands of some
 of the Table1 experiments.}
\end{figure}

    Polarization patterns of ISM clouds are also studied for what
concerning star-formation process for the full understanding of
which, a deep study of present magnetic fields (which align dust
grains) is necessary. This is for instance what SCUBA does at
James Clerk Maxwell Telescope on Mauna Kea, Hawaii. Greaves et al.
\cite{Greaves2000} have observed polarized thermal emission from
the aligned dust grains in the central region of $M82$. $850\mu m$
polarization image of the star-formation region $W3$ has also been
taken and compared with observation made by Hildebrand et
al.\cite{Hildebrand1984} showing a remarkable similarity. $DR21$
is also a widely studied region both from photometric and
polarimetric point of view \cite{Glenn1997}. Thus, if more then
one observation is taken on a single source, these sources could
become efficient calibrators and sources used to test the
polarimeter efficiency.

\section{Conclusions}
\label{sec:concl}

    The main characteristics of our experiment MITO Pol have been
presented. Problems connected with spurious polarization removing
have been analyzed and a method for non-gaussian atmospheric
fluctuation has been presented. This will be applied on MITO Pol
data and will reduce atmospheric contamination that all
ground-based experiments necessary undergo. We think that our
polarimeter could give a contribution to the still open problem of
the detection of CMB polarization and polarized foregrounds.
    First step of MITO Pol will be the observation of foreground
sources in order to be able to create a catalogue of polarized
sources to allow cross-check between various polarization
experiments and to be useful for specific CMB polarization space
mission as MAP, SPOrt or Planck.

\acknowledgments

The progress obtained by the experiment in the last years, could
not have been done without the PhD thesis of Dr. Giampaolo Pisano
who worked extensively on this project.

\bibliography{spiebb}
\bibliographystyle{spiebib}

\end{document}